\documentclass{article}

\usepackage{PRIMEarxiv}

\usepackage[utf8]{inputenc} 
\usepackage[T1]{fontenc}    
\usepackage{hyperref}       
\usepackage{url}            
\usepackage{booktabs}       
\usepackage{amsfonts}       
\usepackage{nicefrac}       
\usepackage{microtype}      
\usepackage{lipsum}
\usepackage{graphicx}
\graphicspath{{media/}}     

\usepackage{amsmath,amssymb,amsfonts}

\usepackage{algorithm}
\usepackage[noend]{algpseudocode}

\usepackage{textcomp}

\newtheorem{dfn}{Definition}

\newtheorem{fct}{Fact}
\newtheorem{prp}{Proposition}
\newtheorem{lem}{Lemma}
\newtheorem{thm}{Theorem}
\newtheorem{crl}{Corollary}
\newtheorem{rem}{Remark}
\newtheorem{exm}{Example}
\newtheorem{apl}{Application}
\newenvironment{prf}{\paragraph{Proof:}}{\hfill$\square$}
  
\usepackage{xspace}
\usepackage{soul}
\usepackage{mathtools}
\usepackage{makecell}
\usepackage{todonotes}
\usepackage{nicefrac}

\newcommand{\ra}{\rightarrow}
\newcommand{\rra}{\rightrightarrows}

\newcommand{\ie}{\unskip, i.\,e.,\xspace}
\newcommand{\eg}{\unskip, e.\,g.,\xspace}

\newcommand{\sut}{\text{s.\,t.\,}}
\newcommand{\rt}{r.\,t.\xspace}

\newcommand{\nrm}[1]{\left\lVert#1\right\rVert}

\newcommand{\abs}[1]{\left\lvert#1\right\rvert}
\newcommand{\scal}[1]{\left\langle#1\right\rangle}

\newcommand{\N}{\ensuremath{\mathbb{N}}}
\newcommand{\Z}{\ensuremath{\mathbb{Z}}}
\newcommand{\Q}{\ensuremath{\mathbb{Q}}}
\newcommand{\R}{\ensuremath{\mathbb{R}}}

\newcommand{\T}{\ensuremath{\mathbb{T}}}
\newcommand{\X}{\ensuremath{\mathbb{X}}}
\newcommand{\Y}{\ensuremath{\mathbb{Y}}}

\newcommand{\U}{\ensuremath{\mathbb{U}}}

\newcommand{\sm}{\ensuremath{\setminus}}
\newcommand{\set}[1]{\ensuremath{\mathbb{#1}}}

\let\emptyset\varnothing

\newcommand*\diff{\mathop{}\!\mathrm{d}}
\newcommand{\eps}{\ensuremath{\varepsilon}}

\newcommand{\spc}{\ensuremath{\,\,}}

\newcommand{\dom}{\ensuremath{\text{dom}}}
\newcommand{\ramp}{\ensuremath{\text{ramp}}}
\newcommand{\pdiff}[2]{ \ensuremath{ \frac{\partial {#1}}{\partial {#2}} } }

\newcommand{\ball}{\ensuremath{\mathcal B}}

\DeclareMathOperator*{\argmin}{arg\,min}

\definecolor{dgreen}{rgb}{0.0, 0.5, 0.0}

\title{On constructive extractability of measurable selectors of set-valued maps
\thanks{\textit{\underline{Published} in IEEE Transactions on Automatic Control}. 
\textbf{DOI:10.1109/TAC.2020.3025303}} 
}

\author{
  Pavel Osinenko \\
  Chemnitz University of Technology \\
  Chemnitz, Germany \\
  \texttt{p.osinenko@email} \\
   \And
  Stefan Streif \\
  Chemnitz University of Technology \\
  Chemnitz, Germany \\
  \texttt{stefan.streif@etit.tu-chemnitz.de} \\
}

\begin{document}
\maketitle

\begin{abstract}
	This paper investigates the possibility of constructive extraction of measurable selector from set-valued maps which may commonly arise in viability theory, optimal control, discontinuous systems etc. 	
	For instance, existence of solutions to certain differential inclusions, often requires iterative extraction of measurable selectors.
	Next, optimal controls are in general non-unique which naturally leads to an optimal set-valued function.
	Finally, a viable control law can be seen, in general, as a selector.
	It is known that selector theorems are non-constructive and so selectors cannot always be extracted.
	In this work, we analyze under which particular conditions selectors are constructively extractable.
	An algorithm is derived from the theorem and applied in a computational study with a three-wheel robot model.
\end{abstract}

\keywords{Algorithms, approximation, selector}

\section{Introduction} \label{sec:intro}

Selector extraction accompanies many control-theoretic analyses.
For instance, existence of solutions in sliding mode systems \cite{kachroo1999existence,Slotine1991-nonlin-ctrl,Perruquetti2002-slid-mode,azhmyakov2010optimal,levant2016discretization}, optimal control problems \cite{Vinter2010-opt-ctrl,frankowska1984maximum}, including dynamic programming \cite{yuksel2013stochastic,arapostathis2012ergodic}, robust stabilization \cite{ledyaev1999lyapunov} and more \cite{barmish1978measurable,hernandez1990error,ryan1994nonlinear,wu2006optimal,papageorgiou2017sensitivity,possamai2018stochastic,proskurnikov2019lyapunov} are concerned with various selectors -- measurable, continuous, minimal etc.
Measurable selector theorems are absolutely crucial for Filippov solutions \cite{Aubin2012-diff-incl} and viability theory \cite{Aubin2011-viab-thr}.
Despite the prevalence of selection theorems in control, they do not allow one to actually compute a selector \cite{Aubin2011-viab-thr}.
However, selectors might be useful even in such practical tasks like stabilization, as will be demonstrated in Section \ref{sec:comp-study} on an example of a three-wheel robot.
Recently, Tsiotras and Mesbahi \cite{tsiotras2017toward} raised some questions regarding computational matters in what they called a possible precursor to ``algorithmic control theory''.
The current work is somewhat motivated by these concerns and suggests to investigate a possible algorithmic content of a measurable selector theorem, namely, the following, fairly general, one:
\cite[Chapter~XIV,~Theorem~1]{Kuratowski1976-set-thr}, \cite[Chapter~18,~18.13]{Guide2006-inf-dim-analys}:

\begin{thm}[General measurable selector theorem]
	Let $\X$ be a measurable and $\Y$ a separable, completely metrizable topological space ($\R^n$ or any separable Banach space are examples), respectively. Let $F: \X \rra \Y$ be a weakly measurable set-valued function with closed values. Then, $F$ admits a measurable selector. 
	\label{thm:meas-sel-class}
\end{thm}

Here, a \textit{set-valued function} (abbreviated SVF in the following) $F: \X \ra \Y$ is a mapping that maps points in $\X$ into non-empty (also called ``inhabited'') subsets of $\Y$.
A mapping $f: \X \ra \Y$ is called \textit{selector} if, for all $x \in \X$, it holds that $f(x) \in F(x)$.
For the details of weak measurability, which are not essential for the current work, the reader may refer to \cite[Chapter~11]{Guide2006-inf-dim-analys}).

Let us now briefly discuss where selectors may arise (we give a more detailed discussion in Section \ref{sec:applications}).
First, consider the following

\begin{apl}
	For a constrained dynamical system
	\begin{align*}
	\dot x = & f(x, u), x(0) = x_0, \\
	& \sut x \in \mathcal X, u \in \mathcal U,
	\end{align*}
	the viability kernel is defined as $\mathcal V(\mathcal X) = \{ x \in \mathcal X : \exists v: [0, \infty) \ra \mathcal U \spc \sut \forall t > 0 \spc x(t) \in \mathcal X \}$, where $v$ denotes a control policy (see 1.2.1.1 in \cite{Aubin2011-viab-thr}).
	We can define a regulation map $\mathcal R_{\mathcal X}$ as follows \cite[1.2.1.3]{Aubin2011-viab-thr}: $\mathcal R_{\mathcal X}$ maps each state $x$ in $\mathcal X$ into a subset $\mathcal R_{\mathcal X}(x)$ of controls $u \in \mathcal U$ which are viable in the sense that $f(x, u)$ is tangent to $\mathcal X$ at $x$: $\mathcal R_{\mathcal X}(x) := \{ u \in \mathcal U: f(x, u) \in T^{**}_{\mathcal X}(x) \}$.
	Here $T^{**}_{\mathcal X}(x)$ denotes the convexified tangent cone (see \cite[Definitions~11.2.1,18.4.8]{Aubin2011-viab-thr}).
	Suppose that, for all $x \in \mathcal X \sm \mathcal S$, for some closed set $\mathcal S$, the regulation map admits closed non-empty values.
	Then, a particular viable feedback can be found by extracting a measurable selector from $\mathcal R_{\mathcal X}$.
	\label{apl:viab-feedback}
\end{apl}

In optimal control, especially dynamic programming \cite{denardo1967contraction}, and economical optimization, the Berge's maximum theorem \cite{Berge1963-top-spcs} plays the central role.
Consider the following

\begin{apl}
	Let $J$ be a continuous function from a product metric space $\X \times \U$ to $\R$, and $R: \X \rra \U$ be a compact-valued continuous SVF.
	By the maximum theorem, the function $J^*(x) := \max_{u \in R(x)} J(x,u)$ is continuous, and the SVF $U^*(x) = \{ u \in R(x) : J(x,u) = J^*(x)\}$ is upper hemi-continuous.
	A particular optimal control (or policy) can be found by a (in general, measurable) selector from $U^*$.
	\label{apl:opt-ctrl}
\end{apl}

In this work, we approach Theorem \ref{thm:meas-sel-class} in the setting of \cite{Bishop1967-constr-analys,Bishop1985-constr-analysis} to investigate a new and constructive version of Theorem \ref{thm:meas-sel-class}.
The goal is to find out under what conditions selectors can be extracted up to any finite arithmetic precision and to suggest a corresponding algorithm.
The setting of \cite{Bishop1967-constr-analys,Bishop1985-constr-analysis} accounts for finite arithmetic precision by removing decision rules of the kind $x \le a \lor x \ge a$ where $x,a$ are any real numbers.
Recently, it was shown how finite arithmetic precision, leading to inexact optimization, might compromise practical stabilization methods \cite{Osinenko2018-pract-stabilization}.
In the following, it is demonstrated that the above decision rule poses specific problems to measurable selector extractability.
Essential for the proof of Theorem \ref{thm:meas-sel-class} is finding so-called \textit{countable reductions} of sequences of sets.
For a sequence of sets $\{ X_n \}_n$, a countable reduction is a sequence $\{ K_n \}_n$ of mutually disjoint sets \sut each $K_n$ is in the corresponding $X_n$ and $\bigcup_n K_n = \bigcup_n X_n$ (whenever we write $\bullet_n$, where $\bullet$ denotes a set or an algebraic operation, the index $n$ runs through all $\N$).
Roughly speaking, countable reduction makes from an arbitrary collection of sets another collection, which covers the original one, but with all the overlapping is removed.
Non-constructivity of the countable reductions can be demonstrated in the following example.

\begin{exm}
	Let $\{X_n\}_n$ be a sequence of intervals in $\R$ defined as follows:
	\begin{equation}
	X_n = \begin{cases}
	[0, \frac{1}{2}], & n \text{ is odd}, \\
	[\frac{1}{4}, 1], & n \text{ is even}.
	\end{cases}
	\label{eqn:Brouwer-exm}
	\end{equation}
	We have $\bigcup_n X_n = [0, 1]$.
	Suppose there exists a countable reduction $\{ K_n \}_n$ \sut $\bigcup_n K_n = \bigcup_n X_n = [0, 1]$. Then, $\{ K_n \}_n$ must have the following form:
	\begin{equation}
	K_n = \begin{cases}
	[0, a], & n \text{ is odd}, \\
	(a, 1], & n \text{ is even},
	\end{cases}
	\label{eqn:Brouwer-exm-count-red}
	\end{equation}	
	where $a \in [\frac{1}{4}, \frac{1}{2}]$.
	The assumption $\bigcup_n K_n = \bigcup_n X_n = [0, 1]$ implies that, for any real number $x$ in the unit interval, it is decidable that $x \le a \lor x \ge a$.
	Subsequently, extracting selectors in FPA, if an algorithm involves a decision of the kind $x \le a \lor x \ge a$, numerical uncertainty may arise.
	The consequence of this is that measurable selectors cannot always be effectively found, as pointed out in \cite{Aubin2011-viab-thr}.
	\label{exm:count-red}
\end{exm}

Examples like the one above demonstrate possible difficulties that may arise in applications due to non-constructivity of underlying derivations.
We have, evidently, $\bigcup_n K_n \subseteq \bigcup_n X_n$, but $\bigcup_n X_n \not\subseteq \bigcup_n K_n$.
What can still be proven is that $\bigcup_n K_n$ is dense in $\bigcup_n X_n$ (see Proposition \ref{prp:count-red-dense} in the appendix).

The central milestones for deriving a version of Theorem \ref{thm:meas-sel-class} with constructive selector extractability are \emph{representable} domains and SVFs, which are discussed in Section \ref{sec:domains}.
The central result is Theorem \ref{thm:meas-sel} in Section \ref{sec:result} followed by Algorithm \ref{alg:meas-sel-extraction} which allows extracting selectors up to any approximation error (details provided).
Applications are discussed in Section \ref{sec:applications}.
A computational study with a three-wheel robot is given in Section \ref{sec:comp-study}. 

\section{Preliminaries} \label{sec:prelim}

In this section, we provide the essential notions for the forthcoming analysis and recall some basic lemmas.

A sequence of some objects is denoted in the form $\{ \bullet_n \}_n$ to stress that iteration is done through the index $n$. 
The closure of a set $A$ is denoted by $\bar A$.
An open ball centered at $x$ and with radius $r$ is denoted by $\ball(x, r)$.
A ball of radius $r$ centered at the origin is denoted $\ball_r$. 
For a located set $A$, we define the distance as $|x - A| \triangleq \inf_{a \in A} |x -a|$ (the set $A$ is located if this distance exists, not all sets are constructively located -- see \cite{Bridges1996-sets}). 
An open interval $(a, b), a < b$ is called \textit{basic interval}. A set $I$ is called \textit{basic set} if it a basic interval, a singleton, or the empty set $\emptyset$. We define $|X - \emptyset| := \infty$ for any located set $X$. Clearly, every basic set is located.
We consider \textit{generalized basic sets} of the form $\bigcup_i I_i$ where each $I_i$ is a basic set. If each $I_i$ is a basic interval, then $\bigcup_i I_i$ is called \textit{generalized interval}. If $x$ is in a generalized basic set $J = \bigcup_i I_i$, then there is some $I_i$ \sut $x \in I_i$. 
Consider the function $\| \bullet \|$ on $\mathcal P (\R)$ as follows: if $X$ is a singleton, or the empty set, then $\|X\| = 0$, if $X$ is an interval of the form $(a,b)$ or $[a,b]$ with $a < b$, then $\|X\| = b-a$, if $X$ is a generalized basic set of the form $\bigcup_i I_i$, then $\|X\| = \sum_i \|I_i\|$ \ie without accounting for overlapping, which is along the lines of \cite{Ye2011-SF} and, in particular, $\| \bullet \|$ differs to the standard measure.
For a sequence of generalized basic sets $X = \{J_n\}_n$ with $J_n = \bigcup_m I_{nm}$, $\Gamma_X$ denotes the set of endpoints of all $I_{nm}, n,m \in \N$. 
We say that a set $A$ is \textit{well-contained} in another set $B$,  denoted by $A \Subset B$, if there exists $r > 0$ \sut $\forall x \in A \spc \forall y \spc |x-y| \le r \implies y \in B$.
A set $A$ is called \textit{weakly detachable} (cf. \cite[p.~70]{Bishop1985-constr-analysis}) if, for any $q \in \Q$, it holds that $q \in A \lor q \not\in A$.
A \textit{regular $\delta$-mesh} is a sequence of points $\{\frac{n}{\delta}\}_n, n \in \Z$.
A set $A$ is \textit{finite} if there exists a bijection $\{1, \dots, N\} \ra A$ for some $\N$.
If this map is a surjection, then $A$ is \textit{subfinite} (classically, subfinite sets are finite).
Thus, if a set is finite, then it is accompanied by the said bijection as a witness.
An \textit{ordered} finite set $X = \{x_n\}_{n \le N}, N \in \N$ is equipped with the relation $\forall n \spc x_n < x_{n+1}$.
Every subset of an ordered finite set is, again, finite.
A function $\hat f$ is called \textit{piece-wise constant extension} of a map $f$ on some ordered finite set $X = \{x_n\}_{n \le N}, N \in \N$ if $\hat f(x) \equiv f(x_n), \forall x \in [x_n, x_{n+1}], \forall n$.
For a closed basic set $I=[a,b], b \le a$, we denote by $\ramp(I, c, d)$ the ramp function $f: I \ra \R$ \sut $f(a) = c$ and $f(b) = d$.
If $I$ is empty, then $\ramp(I, c, d)$ is the empty function.
In the following, for brevity, we consider $\X = \R$ and $\Y = [0, 1]$, 
whereas the constructions extend straightforwardly to the case $\X = \R^n, \Y = \ball_1 \subset \R^m$ (see Section \ref{sec:result}). 
In the subsequent definitions, functions from $\X$ to $\Y$ are meant.

\begin{dfn}[Measurable function]
	A function $f$ is called measurable if, for any $\eps > 0$, there exists a generalized interval $J$ with $\|J\| \le \eps$ and a continuous function $g$ \sut $|f(x) - g(x)| \le \eps$ for any $x \in \X \sm J$.
	The domain of $f$ is denoted $\dom(f)$.
	\label{dfn:meas-fnc}
\end{dfn}

We will use the types of convergence summarized in the following definitions.

\begin{dfn}[Uniform convergence]
	A sequence $\{ f_n \}_n$ of measurable functions converges to a measurable function $f$ uniformly on a basic interval $I \in \X$ if, for any $\eps > 0$, there exists $N \in \N$ \sut $\forall x \in I \spc \forall n \ge N \spc |f_n(x) - f(x)| < \eps$.
\end{dfn}

\begin{dfn}[Convergence almost uniformly]
	A sequence $\{ f_n \}_n$ of measurable functions converges to a measurable function $f$ almost uniformly if, for any $\eps > 0$ and any basic interval $I \subset \X$, there exists a generalized interval $J$ with $\|J\| < \eps$ \sut $\{f_n\}_n$ converges to $f$ uniformly on $I \setminus J$.
	\label{dfn:conv-almost-unif}
\end{dfn}

\begin{dfn}[Convergence in measure]
	A sequence $\{ f_n \}_n$ of measurable functions is called Cauchy in measure if, for any $\eps > 0$ and any basic interval $I \subset I$, there exists $N \in \N$ \sut for all $n > N$, there exists a generalized interval $J$ with $\|J\| < \eps$ \sut $|f_n(x) - f(x)| < \eps$ for any $x \in I \sm J$.
	\label{dfn:conv-in-meas}
\end{dfn}

\begin{dfn}[Cauchy in measure]
	A sequence $\{ f_n \}_n$ of measurable functions is called Cauchy in measure if, for any $\eps > 0$ and any basic interval $I \subset I$, there exists $N \in \N$ \sut for all $n,m > N$, there exists a generalized interval $J$ with $\|J\| < \eps$ \sut $|f_m(x) - f_n(x)| < \eps$ for any $x \in I \sm J$.
	\label{dfn:Cauchy-in-meas}
\end{dfn}


\begin{lem}[6.36 in \cite{Ye2011-SF}]
	If a sequence $\{ f_n \}_n$ of measurable functions is Cauchy in measure, then there exists a measurable function $f$ \sut $\{ f_n \}_n$ converges to $f$ in measure and a subsequence of $\{ f_n \}_n$ converges to $f$ almost uniformly.
	\label{lem:cauchy-meas-conv}
\end{lem}

We can construct countable reductions as in the standard case, but as pointed out in Section \ref{sec:intro}, countable reductions need not to coincide with the original sets.
Therefore, we only have the following fragment of the countable reduction result:

\begin{lem}[Countable reduction]
	Consider a sequence of generalized basic sets $\{ J_n \}_n$. There exists a sequence of generalized basic sets $\{ K_n \}_n$ \sut $\forall n \spc K_n \subseteq J_n$ and $K_n$ are mutually disjoint and $\bigcup_n K_{n}$ is dense in $\bigcup_n J_{n}$.
	\label{lem:count-red}
\end{lem}
\begin{prf}
	The construction of the generalized basic sets $K_n$ goes along the lines of the original proof of \cite[Theorem~4.5.1]{Kuratowski1976-set-thr}. Define 
	\begin{equation}
		\label{eqn:count-red}
		L_{nm} := I_{nm} \sm \bigcup_{ \substack{ \beta(k, l) < \beta(n,m) } } I_{kl},	
	\end{equation}
	where $\beta$ is a bijection between $\N^2$ and $\N$. Let $n,m,n',m'$ be \sut $\beta(n,m) < \beta(n',m')$ and assume $x \in L_{nm} \land x \in L_{n'm'}$.
	Then, $x$ is in $I_{nm}$.
	On the other hand, $\forall k,l \spc \beta(k,l) < \beta(n',m') \spc x \notin I_{kl}$ and so $x \notin I_{nm}$, a contradiction.
	Therefore, all $L_{nm}$ and, consequently $K_n := \bigcup_m L_{nm}$ are mutually disjoint, respectively.
	The density part is proven in the the appendix, Proposition \ref{prp:count-red-dense}.
	Thus, $\{ K_n \}_n$ is the required countable reduction.
\end{prf}

What Lemma \ref{lem:count-red} roughly says is that we can still algorithmically remove overlapping from a given collection of sets, but the ``covering'' is not provided on a set of measure zero (please refer to Example \ref{exm:count-red}).

In the following, exactly the construction in the proof of Lemma \ref{lem:count-red} will be meant by the countable reduction. Now, we proceed to the essential part of the current work -- representable domains and SVFs.

\section{Representable domains} \label{sec:domains}

In this section, we introduce the essential ingredients for the current analysis. First, some intermediate results are required.

\begin{prp}
	For any generalized basic set $J$, its countable reduction $K$ satisfies $\|J\| - \|K\| = 0$.
	\label{prp:count-red-meas}
\end{prp}
\begin{prf}
	Let $J$ be represented as $J = \bigcup_i I_i$ and its countable reduction as $K = \bigcup_i L_i$. We proceed by induction noticing that $\|\Gamma_J\| = 0$ since $\Gamma_J$ is countable. Since $L_1 = I_1$, it holds that $\|I_1\| - \|L_1\| = 0$. Assume that any interval $I = (a, b), b < a$ contained in $\bigcup_{k=1}^i I_k \sm \Gamma_i$ is in $\bigcup_{k=1}^i L_k$, whence $\| \bigcup_{k=1}^i I_k \sm \Gamma_i \| - \| \bigcup_{k=1}^i L_k \| = 0$, where $\Gamma_i$ is the set of endpoints of $\{I_k\}_{k=1,\dots,i}$. We show that any interval $I = (a, b), b < a$ contained in $\bigcup_{k=1}^{i+1} I_k \sm \Gamma_i$ is in $\bigcup_{k=1}^{i+1} L_k$. First, it holds trivially that $\bigcup_{k=1}^{i+1} L_k \sm \Gamma_{i+1} \subseteq \bigcup_{k=1}^{i+1} L_k \sm \Gamma_i$. Express the latter in the following form:
	\begin{align*}
	\left( \bigcup_{k=1}^i I_k \sm \Gamma_i \right) \bigcup \left( \bigcup_{p \in P} \bigcap_{e \in E_p} I_q \right) \bigcup \left( I_{i+1} \sm \bigcup_{r \in R} \bigcap_{w \in W_r} I_w \right),
	\end{align*}
	where $P, E, R, W_r \subseteq \{0,\dots,n!\}$ and $e, w \in \{0,\dots,n\}$. If $I \subseteq \bigcup_{k=1}^i I_k \sm \Gamma_i$, then, by the induction hypothesis, $I \subseteq \bigcup_{k=1}^i L_k \subseteq \bigcup_{k=1}^{i+1} L_k$. If $I \subseteq \bigcup_{p \in P} \bigcap_{q \in E_p} I_q $, then $I \subseteq I_{i+1}$ and so $I_{i+1}$ is a witness for $I \subseteq \bigcup_{k=1}^{i+1} L_k$. If $I \subseteq \bigcap_{w \in W_r} I_w$ for some $p$, then $I_{\min(E_p)}$ is a witness for $I \subseteq \bigcup_{k=1}^{i+1} L_k$.
\end{prf}

\begin{crl}
	It follows that $J \sm \Gamma_J \subseteq K \sm \Gamma_J \subseteq K$. Also, $\Gamma_K \subseteq \Gamma_J$ since the countable reduction does not introduce any new endpoints.
\end{crl}

\begin{crl}
	For any sequence of generalized basic sets $X = \{J_n\}_n$, its countable reduction $K = \{K_n\}_n$ satisfies $\|X\| - \|K\| = 0$.
\end{crl}

\begin{crl}
	Trivially, for any finite ordered set $X = \{x_n\}_{n \le N}, n \in N$ seen as a union of singletons, its countable reduction $K$ coincides with $X$.
\end{crl}

Finally, the representable domain is introduced:

\begin{dfn}[Representable domain]
	A sequence of generalized basic sets $X$ is called a representable domain relative to (further abbreviated ``\rt'') $\X$ if, for any $\eps > 0$, there exists a generalized interval $M$ with $\|M\| \le \eps$ \sut
	\begin{enumerate}
		\item $\Gamma_X \Subset M \sm \Gamma_M$,
		\item $\X \sm M \subseteq X$. 
	\end{enumerate}
	\label{dfn:represent-dom}
\end{dfn}

The idea behind Definition \ref{dfn:represent-dom} is that it entails an algorithm which gives an ``arbitrarily thin'' generalized interval such that the representable domain covers the total space minus this generalized interval.

\begin{rem}
	It holds that if $X$ is a representable domain \rt $\X$, then $\|X\| - \|\X\| = 0$.  
\end{rem}

We sometimes specify explicitly which $\eps > 0$ the witness for representability is related to and denote it $M(\eps)$.

\begin{rem}
	If $X$ is a representable domain, then, for any $\eps>0$, there exists also a generalized basic set $M'$ with the properties listed in Definition \ref{dfn:represent-dom} and, additionally, $M'$ consists of disjoint basic sets. This follows by calculating the countable reduction $M'$ of $M(\eps)$ and applying the fact that $M \sm \Gamma_M \subseteq M' \sm \Gamma_M$ from Proposition \ref{prp:count-red-meas}.
	\label{rem:disjoint-repres-witness}
\end{rem}

An important property of representable domains consisting of mutually disjoint sets is that piece-wise constant maps on them are measurable:

\begin{fct}
	Let $X = \bigcup_n J_n$ be a representable domain \rt $\X$ \sut all $J_n, n \in \N$ are mutually disjoint. Let $f$ be a mapping that takes constant values on $J_n, n \in \N$. Then, $f$ is a measurable function.
	\label{fct:meas-fnc-repres}
\end{fct}
\begin{prf}
	Fix an $\eps>0$ and let $M$ be the witness for representability of $X$ \rt $\X$. Let $M' = \bigcup_i V'_i$ be its countable reduction. Define a mapping $g$ as follows. If $x \in \X \sm M$, set $g(x) := f(x)$. If $x \in M'$, then $\exists i \spc x \in V'_i$. Define $g(x) := \ramp(\bar V'_i, f_1, f_2)(x)$ where $f_1, f_2$ are the values of $f$ at the left and right endpoints of $\bar V'_i$, respectively. Since $M' \subseteq M$, it holds that $\X \sm M \subseteq \X \sm M'$, whence, trivially, $\X \sm M'$ is dense in $\X \sm M$. By Proposition \ref{prp:count-red-dense}, $M'$ is dense in $M$. Therefore, $\X \sm M' \bigcup M'$ is dense in $\X \sm M \bigcup M$. On the other hand, $\X \sm M \bigcup M \subseteq \Gamma_M$ and $\Gamma_M$ is countable. Applying Theorem 2.19 from \cite{Bishop1985-constr-analysis}, conclude that $\X \sm M \bigcup M$ is dense in $\X$. Therefore, $g$ is defined on a dense subset of $\X$. Apply Lemma 3.7 from \cite{Bishop1985-constr-analysis} to extend $g$ to a continuous function on $\X$. Clearly, $g(x) = f(x)$ on $\X \sm M'$, and, thus, on $\X \sm M$ which implies $f$ is measurable.
\end{prf}

Some properties of representable domains are summarized in the following proposition.

\begin{prp} The following properties of representable domains hold.
	\begin{enumerate}
		\item The countable reduction $K$ of a representable domain $X$ \rt $\X$ is a representable domain \rt $\X$, \label{enum:represent-dom-prop1}
		\item For any two representable domains $X_1, X_2$ \rt $\X$, the intersection $X' = X_1 \bigcap X_2$ is a representable domain \rt $\X$, \label{enum:represent-dom-prop2}
		\item For any two representable domains $X_1 = \bigcup_n J_{1,n}$, $X_2 = \bigcup_n J_{2,n}$ \rt $\X$, the term-wise intersection $\tilde X = \bigcup_n J_{1,n} \bigcap J_{2,n}$ is a representable domain if $X_1 \subseteq \tilde X$. \label{enum:represent-dom-prop3}
	\end{enumerate}
	\label{prp:represent-dom-props}
\end{prp} 
\begin{prf}
For \ref{enum:represent-dom-prop1}, fix any $\eps > 0$ and let $M$ be as in Definition \ref{dfn:represent-dom}. We have:
\begin{align*}
& \X \sm M \subseteq X \sm \Gamma_X \implies \\
& \X \sm M \subseteq K & (X \sm \Gamma_X \subseteq K \text{ by Proposition \ref{prp:count-red-meas}} ).
\end{align*}
Therefore, a witness $M(\eps)$ for representability of $X$ \rt $\X$ is directly a witness for representability of $K$ \rt $\X$.

Now, property \ref{enum:represent-dom-prop2}. Let $M_1(\eps), M_2(\eps)$ be the witnesses for representability of $X_1, X_2$ \rt $\X$ for some $\eps > 0$, respectively. It holds that:
\begin{align*}
& \X \sm M_1 \subseteq X_1 \land \X \sm M_2 \subseteq X_2 \implies \\
& \X \sm \left( M_1 \bigcup M_2 \right) \subseteq X_1 \land \X \sm \left( M_1 \bigcup M_2 \right) \subseteq X_2 \implies \\
& \X \sm \left( M_1 \bigcup M_2 \right) \subseteq X_1 \bigcap X_2 = X'.
\end{align*}
Define $M' := M_1\left( \frac{\eps}{2} \right) \bigcup M_2\left( \frac{\eps}{2} \right)$.
Clearly, $\Gamma_{X'} \Subset M'$. Therefore, $M'$ is a witness for representability of $X'$ \rt $\X$.

Lastly, consider property \ref{enum:represent-dom-prop3}. Let $\tilde M := M_1\left( \frac{\eps}{2} \right) \bigcup M_2\left( \frac{\eps}{2} \right)$ for some fixed $\eps>0$, and so $\|\tilde M\| \le \eps$.
Then, $\Gamma_{\tilde X} \subseteq \Gamma_{X_1} \bigcup \Gamma_{X_2} \Subset \tilde M$.
Further,
\begin{align*}
& \X \sm M_1 \subseteq X_1 \subseteq \tilde X, \\
& M_1 \subseteq \tilde M \implies \X \sm \tilde M \subseteq \X \sm M_1 \implies \\
& \X \sm \tilde M \subseteq \tilde X.
\end{align*}
\end{prf}

Finally, we may introduce the type of SVFs used in the core of the current work in Section \ref{sec:result}.

\begin{dfn}[Representable SVF]
	A SVF $F: \X \rra \Y$ is called representable if, for any $x \in X$, $F(x)$ is closed and located in the sense of \cite{Bishop1985-constr-analysis} \ie that $|\bullet - F(x)|$ be computable (see also discussion on computability in Example \ref{exm:count-red}), and for any finite sequence $\{r_i\}_i^N \subset \Y$ and $\delta \in \Q_{>0}$, there exist sets
	\[
	C_i^{\delta} = \{ x \in \X : | r_i - F(x) | \le \delta \}
	\]
	and $\bigcup_{i \le N} C_i^{\delta}$ is a representable domain \rt $\X$.
	\label{dfn:represent-mfnc}
\end{dfn}

\begin{exm}
	A simple example of a representable SVF $F: \X \rra \Y$ is a piece-wise finite union of weakly detachable basic sets in the following form:
	\begin{equation}
	\begin{aligned}
	& F(x) = \sum_{k=1}^n \chi_{[a_k, b_k]}(t) \bigcup_{j \in J_k} A_j, \spc \forall k \spc J_k \simeq \{1, \dots, N_k\}, \\
	& \forall k \spc a_k \le b_k \le a_{k+1},
	\end{aligned}
	\label{eqn:exm-represent-multifnc}
	\end{equation}
	where $\chi_{[a_k, b_k]}$ is the characteristic function of $[a_k, b_k]$. In this case, each $C_i^{\delta}$ is either $\emptyset$ or a finite union of some of the intervals $[a_k, b_k], k \in \{1, \dots, n\}$ and there are finitely many mutually different $C_i^{\delta}$. Therefore, $\Gamma_{\bigcup_i C_i^{\delta}}$ is subfinite and we can always find an $M$ as per Definition \ref{dfn:represent-dom}.
\end{exm}

\section{Selector theorem with constructive extractability} \label{sec:result}

The following theorem is a version of the one by \cite{Kuratowski1976-set-thr} with constructively extractable selectors.

\begin{thm}
	Let $F: \X \rightrightarrows \Y$ be a representable SVF. Then, $F$ admits a measurable selector.
	\label{thm:meas-sel}
\end{thm}
\begin{prf}
We construct inductively a sequence of measurable functions $f_n$ to $\Y$ with representable domains \rt $\X$ satisfying:
\begin{enumerate}
	\item $|f_{n}(x) - F(x)| < \frac{1}{2^n}, \forall x \in \dom(f_n),$
	\item $\dom(f_{n-1}) = \bigcup_{i \le N_{n-1}} Q^{n-1}_i, N_{n-1} \in \N$ is representable \rt $\X$,
	\item $Q^{n-1}_i, i \le N_{n-1}$ are mutually disjoint and $f_{n-1}$ takes constant values on these sets,
	\item $\dom(f_n) \subseteq \dom(f_{n-1})$,
	\item $|f_n(x) - f_{n-1}(x)| < \frac{1}{2^{n-1}}, \forall x \in \dom(f_n).$
\end{enumerate}
Define $f_1(x) \equiv 0$.
Let $\{r^n_i\}_{i \le N_n}$ be a regular $\frac{1}{2^{n+1}}$-mesh on $\Y$.
Consider the following sets:
\begin{equation}
	\begin{aligned}
		C^n_i & := \left\{ x \in \X : | r^n_i - F(x) | < \frac{1}{2^n} \right\}, \\
		D^n_i & := \left\{ x \in \dom(f_{n-1}) : | r^n_i - f_{n-1}(x) | < \frac{1}{2^{n-1}} \right\}, \\
		A^n_i & := C^n_i \bigcap D^n_i.
	\end{aligned}
\label{eqn:meas-sel-sets}
\end{equation}
Since $D^2_i, i \le N_n$ are either $\emptyset$ or $\X$, $\bigcup_{i \le N_n} D^2_i$ is a representable domain \rt $\X$. Suppose, $f_{n-1}$ has been constructed. It follows that $\bigcup_{i \le N_n} D^n_i$ is a representable domain \rt $\X$ since each $D^n_i$ is effectively a union of some of the sets $Q^{n-1}_i$. Let $x \in \dom(f_{n-1})$. By the induction hypothesis, there exists $y \in F(x)$ \sut $|y - f_{n-1}(x)| < \frac{1}{2^{n-1}}$. 
Since there always exists a ball of radius $\frac{1}{2^{n+1}}$ in $\ball(f_{n-1}(x), \frac{1}{2^{n-1}}) \bigcap \ball(y, \frac{1}{2^n})$, we can find an $r^n_i$ \sut
\begin{equation}
\begin{aligned}
& |r^n_i - f_{n-1}(x)| < \frac{1}{2^{n-1}} & & \land & & |r^n_i - y| < \frac{1}{2^n}.
\end{aligned}
\label{eqn:meas-sel-cond}
\end{equation}
Therefore, $x \in A_i$. Since $\bigcup_{i \le N_n} D^n_i \subseteq \dom(f_{n-1}(x))$, the same argument holds for any $x \in \bigcup_i D^n_i$ and, therefore, $\bigcup_{i \le N_n} D^n_i \subseteq \bigcup_{i \le N_n} A^n_i$.
Since $F$ is representable, $\bigcup_{i \le N_n} C^n_i$ is a representable domain \rt $\X$.
Apply Proposition \ref{prp:represent-dom-props} to conclude that $\bigcup_{i \le N_n} A^n_i$ is a representable domain \rt $\X$. 
Let $\bigcup_{i \le N_n} Q^n_i$ be the countable reduction of $\bigcup_{i \le N_n} A^n_i$. By Proposition \ref{prp:count-red-meas},
\begin{equation}
\begin{aligned}
\| \bigcup_{i \le N_n} Q^n_i \| - \| \bigcup_{i \le N_n} A^n_i \| = 0, \\
\| \bigcup_{i \le N_n} Q^n_i \| - \|\X\| = 0.
\end{aligned}
\label{eqn:meas-sel-meas-zero}
\end{equation}
By Proposition \ref{prp:represent-dom-props}, $\bigcup_{i \le N_n} Q^n_i$ is a representable domain \rt $\X$.
Define $\dom(f_n) : = \bigcup_{i \le N_n} Q^n_i$ and $f(x) :\equiv r^n_i \iff x \in Q^n_i$.
Clearly, $\dom(f_n) \subseteq \dom(f_{n-1})$. Due to Fact \ref{fct:meas-fnc-repres}, $f_n$ is a measurable function. We have $x \in Q^n_i \implies x \in C^n_i$ and, thus, $|f_n(x) - F(x)| < \frac{1}{2^n}$. On the other hand, $x \in Q^n_i \implies x \in D^n_i$ and so $| f_n(x) - f_{n-1}(x) | < \frac{1}{2^{n-1}}$. Therefore, $f_n$ is constructed as required.

Fix an arbitrary $\eps > 0$ and choose $N$ satisfying $\nicefrac{1}{2^{N-1}} < \eps$. Fix some $n, m > N$. Let $M_n(\frac{\eps}{2})$ and $M_m(\frac{\eps}{2})$ be the witnesses for representability \rt $\X$ of $\dom(f_n)$ and $\dom(f_m)$, respectively. Set $M := M_n(\frac{\eps}{2}) \bigcup M_m(\frac{\eps}{2})$. Then, $\|M\| < \eps$ and $|f_n(x) - f_m(x)| < \eps, \forall x \in \X \sm M$. Thus, $\{f_n\}_n$ is Cauchy in measure. Consequently, by Lemma \ref{lem:cauchy-meas-conv}, $\{f_n\}_n$ converges to a measurable function $f$ in measure and a subsequence of $\{f_n\}_n$ converges to $f$ almost uniformly. Since $\forall x \in X \spc F(x)$ is closed, it follows that $\forall x \in X \spc f(x) \in F(x)$.
\end{prf}

\begin{rem}
	Notice, we do not consider a dense sequence in $\Y$, but regular finite meshes, which are refined for each $n$.
	Thus, this crucial step in the proof uses just finite sequences.
\end{rem}

A particular algorithm can be extracted from Theorem \ref{thm:meas-sel} which yields a measurable selector for the given representable SVF up to any prescribed approximation error in the sense of the condition 1) of the proof of Theorem \ref{thm:meas-sel} (see Algorithm \ref{alg:meas-sel-extraction}).

\begin{algorithm}[H]
	\caption{Measurable selector extraction}
	\begin{algorithmic}
		\Require Representable SVF $F: \X \rightrightarrows \Y$, approximation error $\eps = \frac{1}{2^n}$
		\State\phantom{b}\vspace{-10pt}
		\State Initialize $f_1 :\equiv 0$
		\For {$k \in \{2,\dots,n\}$}
			\State Generate $\{r^k_i\}_{i \le N_k}$, a regular $\frac{1}{2^{k+1}}$-mesh on $\Y$
			\State Compute sets $C^k_i, D^k_i, A^k_i$ as per \eqref{eqn:meas-sel-sets}
			\State Compute $\{Q^k_i\}_{i \le N_k} := \text{countableRed}\left( \{A^k_i\}_{i \le N_k} \right)$ by \eqref{eqn:count-red}
			\State Set $\dom(f_k) : = \bigcup_{i \le N_k} Q^k_i$ and $f_k :\equiv r^k_i$ on $Q^k_i$.
		\EndFor
		\Return $f_n$, measurable selector of $F$ up to approximation error $\eps$
	\end{algorithmic}
	\label{alg:meas-sel-extraction}	
\end{algorithm}

\begin{crl}
	Assume $\X$ to be compact and let $F$ satisfy the following \emph{weak continuity} condition on $\X$:
	
	\medskip
	
	\noindent for any $\eps > 0$, there exists $\delta > 0$ \sut for any finite ordered set $X = \{x_k\}_{k \le N}, n \in N$ with $\forall k, l \le N \spc |x_k - x_l | < \delta$ and any map $f: X \ra \Y$ satisfying $\forall x \in X \spc |f(x) - F(x)| < \frac{\eps}{2}$, there exists a generalized interval $J$ with $\|J\| < \eps$ \sut the piece-wise constant extension $\hat f$ of $f$ satisfies
	\begin{equation}
	\forall x \in \X \sm J \spc | \hat f(x) - F(x) | < \eps.
	\label{eqn:weak-cont-cond}
	\end{equation}
	Then, $F$ admits a measurable selector.
\end{crl}
\begin{prf}
We construct, similarly to the proof of Theorem \ref{thm:meas-sel}, a sequence of measurable functions $\{\hat f_n\}_n$ and show that it is Cauchy in measure. Fix any $n < m \in \N$. Define the following set:
\begin{equation}
\begin{aligned}
\Gamma_C := \bigcup_{l \le m} \bigcup_{i \le N_m} C^l_i,
\end{aligned}
\label{eqn:meas-sel-crl-endpts}
\end{equation}
where $C^l_i$ are as in \eqref{eqn:meas-sel-sets}.	
Let $\delta$ be a witness for weak continuity of $F$ for $\eps = \frac{1}{2^{n-1}}$. Since $\Gamma_C$ is countable, we can find a finite ordered sequence $X$ with $\forall x_p, x_q \in X \spc |x_p - x_q| < \delta$ \sut $X_n$ is apart from $\Gamma_C$. In this case, proceeding as in the proof of Theorem \ref{thm:meas-sel}, but for $X$ in place of $\X$, we obtain two maps $f_n$ and $f_m$ with $\dom(f_n) = \dom(f_m) = X$ and satisfying:
\begin{equation}
\begin{aligned}
\forall x \in X \spc & |f_n(x) - F(x)| < \frac{1}{2^n} \spc \land |f_m(x) - F(x)| < \frac{1}{2^m}.		
\end{aligned}
\label{eqn:meas-sel-crl-cond}
\end{equation}
Since $F$ is weakly continuous, the piece-wise constant extensions $\hat f_n$ and $\hat f_m$ of $f_n$ and $f_m$, respectively, satisfy
\begin{equation}
\begin{aligned}
& |\hat f_n(x) - F(x)| < \frac{1}{2^{n-1}} \spc \land |\hat f_m(x) - F(x)| < \frac{1}{2^{n-1}} \spc \land \\
& |\hat f_n(x) - \hat f_m(x)| < \frac{1}{2^{n-2}}
\end{aligned}
\label{eqn:meas-sel-crl-extens}
\end{equation}	
on $X \sm J$ for some generalized interval $J$ with $\|J\| < \frac{1}{2^{n-1}}$. Clearly, $f_n$ and $f_m$ are measurable. Thus, $\{f_n\}_n$ is Cauchy in measure. The rest follows.
\end{prf}


The derived results
generalize to the case of SVFs from $\R^{\alpha}$ to $\R^{\beta}, \spc \alpha, \beta \in \N$ as follows.
The distance $|\bullet - \bullet|$ is considered as simply the Euclidean distance.

Basic and generalized intervals are substituted with basic and generalized boxes (hyper-rectangles) determined by the vertices $ \mathcal H = \{ x_l \}_{l=1}^L, x_l \in \R^{\gamma}$, where $L$ is according to the dimension $\gamma$, and $\bigcup_i \mathcal H_i$, where each $\mathcal H_i$ is a basic box, respectively.
Instead of boxes, balls could be used, but boxes are computationally more suitable.
The function $\|\bullet\|$ on $\R^{\gamma}$ is considered as follows: for a basic box, it equals its geometric volume; for $\emptyset$, a singleton and any box boundary, it equals zero.
The sets $\Gamma_X$ where $X$ is a generalized basic set are now the unions of the boundaries of the basic sets constituting $X$.
Generalized intervals $J$ in the definitions of measurable functions and convergence (see Definitions \ref{dfn:meas-fnc}, \ref{dfn:conv-almost-unif}, \ref{dfn:conv-in-meas}, \ref{dfn:Cauchy-in-meas}) are simply substituted with generalized boxes.

Some results, specifically, Lemma \ref{lem:count-red}, Proposition \ref{prp:count-red-meas}, Proposition \ref{prp:count-red-dense} and also the properties of representable domains in Proposition \ref{prp:represent-dom-props} can be generalized straightforwardly noticing that $\|\Gamma_X\| = 0$ for any generalized basic set $X$ since $\Gamma_X$ is a countable union of boundaries of boxes.
In Fact \ref{fct:meas-fnc-repres}, the set $\X \sm M \bigcup M$ is not, in general, dense in $\X$.
It is, however, if $M$ is located \cite{Bridges1996-sets}.
Therefore, we require additionally $M$ in Definition \ref{dfn:represent-dom} to be located.
If $M'$ is the countable reduction of $M$, then it is dense in $M$ and, thus, located as well with the same distance function $|\bullet - M'| = |\bullet - M|$.

Furthermore, the ramp functions on single basic sets $\bar V'_i$ are generalized.
We need to require each $J_n$ in Fact \ref{fct:meas-fnc-repres} to be measurable \ie to have a measurable characteristic function.
In this case, knowing that, for any $\eps$, it holds that $\Gamma_X \Subset M(\eps) \sm \Gamma_M$, there is some box $\mathcal H$ \sut $\Gamma_X + \mathcal H \subseteq M \sm \Gamma_M$.
Therefore, we take witnesses $g_n$ for measurability of $J_n$ \sut $y_n \cdot g_n(x) = f(x)$ on $J_n \sm L_n$, where $L_n$ is a generalized box with $\|L_n\| < \frac{r}{2}$ and $y_n$ is the value of $f$ on $J_n$.
Let, additionally, $X = \bigcup_n J_n$ satisfy the following \emph{weak finite adjacency condition}:

\medskip
\noindent For any basic box $\mathcal H_{x, \delta}$ with a vertex at $x$ and maximum edge length $\delta$, there is a finite set $P_{x, \delta} \subset \N$ of indices \sut $\mathcal H_{x, \delta} \subset \bigcup_{n \in P_{x, \delta}} (J_n + \mathcal C_\delta) \sm \bigcup_{m \in \N \sm P_{x, \delta}} (J_m + \mathcal C_\delta) $, where $\mathcal C_\delta$ is the hypercube centered at the origin with the side length $\delta$.
\medskip

\noindent If $X$ is weakly finitely adjacent, let $\bigcup_m \mathcal H_k$ be \sut \\ $\bigcup_k \mathcal H_k = \R^{\gamma}$ and, for any $x$, there is $k$ \sut $x \in \mathcal H_k$.
Define the required continuous function as $g(x) := \sum_{n \in P_{x_k, r}} y_n \cdot g_n$.
The idea is that any $x$ is in some box of the said cover of $R^\gamma$.
Each box is certainly within $\bigcup_n J_n + \mathcal C_\delta$, but we want it finite, so we select only some suitable finite set of $\bigcup_n J_n + \mathcal C_\delta$ and sum up their respective $g_n$s.
The algorithm entailed in the definition of weak finite adjacency does not lead to a discontinuous $g$ because it may switch the index sets carefully -- include indices whenever there is no certainty that they can be dropped.
At some point, some indices become unnecessary with certainty.
We effectively use the overlap property here.
It is easy to show that weak finite adjacency condition is preserved under countable reduction and union, term-wise intersection and selection of the type $\bigcup_{e \in E} J_n, E \subset \N$.
Finally, recall Lemma \ref{lem:count-red}: if each $I_{nm}$ is measurable then each $L_{nm}$ is measurable also, specifically, with the characteristic function $\chi_{L_{nm}} = \chi_{I_{nm}} - \sum_{\beta(k,l) < \beta(n,m)} \chi_{I_{kl}}$.
Assuming $\bigcup_{i \le N} C^{\delta}_i$ in Definition \ref{dfn:represent-mfnc} are representable \rt $\X$ with located representability witnesses and satisfy weak finite adjacency condition generalizes the constructions in Theorem \ref{thm:meas-sel} apply to the case when $\X = \R^{\alpha}$ and $\Y=[0, 1]^{\beta}, \spc \alpha, \beta \in \N$.
In the next section, we proceed to particular applications of the selector theorem.

\section{Applications} \label{sec:applications}

In this section, we discuss some mathematical results, relevant for control systems, which rely on measurable selector extraction.
The first group of applications deals with dynamical systems described by differential inclusions (DIs) of the form:
\begin{equation}
\label{eqn:diff-incl}
	\dot x \in F(t, x), x(0) = x_0.
\end{equation}
A (local) Filippov solution to \eqref{eqn:diff-incl} is an absolutely continuous function whose derivative satisfies the differential inclusion almost everywhere.
Existence of Filippov solutions depends on the properties of the SVF $F$ such as semi-continuity, image closeness or convexity.
For instance, a common setting of sliding mode control of a system $\dot x = f(x, u)$, where $u \in \U_1 \cup \U_2$, a disconnected compact set, and where the state space is divided into two regions $\X_1, \X_2$ divided by a sliding surface $\Sigma$, is usually treated in the sense of a DI \cite{Perruquetti2002-slid-mode,Slotine1991-nonlin-ctrl}:
\begin{equation}
\label{eqn:SMC-DI}
		\dot x	\in F(x,u) = \begin{cases}
		f_1(x, u), & x	\in \X_1,\\
			(1 - \alpha)f1(x, u) + \alpha f_2(x,u), \alpha \in [0, 1], & x \in \Sigma, \\
		f_2(x, u), & x	\in \X_2,
	\end{cases}
\end{equation} 
where $f_1, f_2$ describe the dynamics of the system on $\X_1 \cup \Sigma \times \U_1$ and $\X_2 \cup \Sigma \times \U_2$, respectively.
The above is a particular example of Filippov regularization of a differential equation with a discontinuous right-hand side.
The DI \eqref{eqn:SMC-DI} is upper semi-continuous, has closed and convex images and thus admits a Filippov solution \cite{Aubin2012-diff-incl,zabczyk2009mathematical}.
A number of existence theorems in Chapter 2 of \cite{Aubin2012-diff-incl} deal with the cases when the DI is: (a) lower or upper semi-continuous with closed, convex images; (b) Lipschitzean; (c) lower semi-continuous with compact (not necessarily convex) images.
The sliding-mode control in the form \eqref{eqn:SMC-DI} thus falls into the category (a).
The category (c) is fairly general, whereas for Lipschitzean DIs, there are applications that study \eg their reachable sets \cite{wolenski1990exponential}, asymptotic stability properties \cite{angeli2004uniform}, and discretized DIs, like those arising in Euler schemes \cite{acary2008numerical}. 
Crucial to the most constructions of solutions is extraction of measurable selectors from the maps of the type $F(\bullet, \xi( \bullet ) )$, where $\xi$ is absolutely continuous, which may arise in an iterative procedure (in the appendix, we give some more details on how it works in case of Lipschitzean DIs).
Therefore, if one is concerned with a verified computation, Theorem \ref{thm:meas-sel} and the respective Algorithm \ref{alg:meas-sel-extraction} may be applied if $F$ is also representable.
In Section \ref{sec:comp-study}, an example is demonstrated, although for the case of practical stabilization, a similar procedure may be applied to Filippov solutions and other problems which we discuss further. 

Now, consider, as the next one, the viability kernel of Application \ref{apl:viab-feedback}.
If the corresponding regulation map $\mathcal R_{\mathcal X}$ happens to be representable, we can apply Theorem \ref{thm:meas-sel} to find a feedback constructively.
Under the condition that $f(\bullet, \mathcal U)$ is a Marchaud SVF, which, in particular, means that the values $f(x, \mathcal U)$ are non-empty, closed and convex, there even exists a continuous selector.
Again, the corresponding result is \emph{per se} not constructive \cite{Aubin2011-viab-thr}.

Finally, recall the Berge's maximum theorem, a central result used in dynamic programming \cite{denardo1967contraction}.
Again, if $U^*$ is representable, we can find an optimal control law constructively by applying Theorem \ref{thm:meas-sel}.
As such, the Berge's maximum theorem is, however, not constructive.
But, under special conditions on maxima, it can be proven constructively \cite{Tanaka2012-maximum-thm}.
Even if these conditions do not hold, we can still construct an approximately optimal SVF $U_{\eps}^*(x) = \{ u \in R(x) : | J(x,u) - J^*(x) | < \eps \}$ for any $\eps$.
For a constructive analysis of continuous approximately optimal control, the reader may refer to \cite{Osinenko2018-constr-aEVTfnc,Osinenko2018-constr-aEVTfnc-Euclid}.
In general, it is important, however, to consider relaxed \ie measurable, controls \cite{filippov1962certain}.

\section{Computational study} \label{sec:comp-study}

This section presents an application of Algorithm \ref{alg:meas-sel-extraction} to the problem of practical stabilization of a three-wheel robot by means of a technique which uses so-called \textit{disassembled subdifferentials}.
First, the model reads:
\begin{equation}
	\label{eqn:NI}
	\tag {NI}
	\dot x = \underbrace{\begin{pmatrix}
		1 \\ 0 \\ -x_2
		\end{pmatrix}}_{=: g_1(x)} u_1 + \underbrace{\begin{pmatrix}
		0 \\ 1 \\ x_1
		\end{pmatrix}}_{=: g_2(x)} u_2.
\end{equation}
As there is no smooth control Lyapunov function (CLF) for \eqref{eqn:NI} \cite{Brockett1983-stabilization}, a variety of non-smooth CLFs and discontinuous stabilization controls were and are still being suggested (see \eg a nice recent numerical study \cite{Braun2017-SH-stabilization-Dini-aim}).
Here, we pick the disassembled subdifferential technique by \cite{Kimura2013-control,Kimura2015-asymptotic} due to its relative simplicity and relevance of selector extraction.
The essential building blocks of this technique are: (a) a disassembled CLF; (b) disassembled subdifferentials.
The idea of the method goes as follows.
Given a (not necessarily non-smooth) proper, positive-definite, locally semiconcave function $V: \R^n \ra \R$, we are interested in filling its non-existing gradients with more general objects (subgradients).
Here, \textit{locally semiconcave} means for any $x, y$ it holds that $V(x) + V(y) - 2 V \left( \frac{x + y}{2} \right) \leq C \nrm{x - y}^2$ for some $C \ge 0$.
This property is ubiquitous in CLFs \cite{Clarke2011-discont-stabilization}.
There are many types of subgradients \cite{Clarke2008-nonsmooth-analys}, but in the disassembled setting, we first formally write down $V$ as
\begin{equation}
	\label{eqn:as-marg-fnc}
	V(x) = \min_{\theta \in \Theta} F(x, \theta),
\end{equation}
where $F$ is a so-called \textit{marginal function} \cite{Cannarsa2004-semiconcave}, and $\Theta$ is a compact set.
Marginal functions in this context are picked smooth and the idea is to take its gradients as subgradients of $V$ by picking the respective minimizers $\theta$ in \eqref{eqn:as-marg-fnc}.
Namely, consider
\begin{dfn}[$F$-disassembled subdifferential]
	\label{dfn:disass-diff}
	Let $V: \X \ra \R, \spc \X \subseteq \R^n$ be a locally semiconcave function and let $F: \set K \times \Theta \ra \R$ be a marginal function, where $\set K \subset \R^n$ and $\Theta \subset \R^{2n}$ are compact sets.
	The SVF $\partial_D^F V: \X \ra \R^n$ given as
	\begin{equation} \label{eqn:disass-diff}
		\partial_D^F V(x) \triangleq \left\{ \pdiff{F(x, \theta^*)}{x}: \theta^* \in \argmin_{\theta \in \Theta} F(x;\theta) \right\}
	\end{equation}
	is called \textit{$F$-disassembled subdifferential}.
	A single element of $\partial_D^F V(x)$ is called \textit{$F$-disassembled subgradient} and is in general a measurable function of $x$.
\end{dfn}
It should be noted that the minimizers $\theta^*$ above need not to be unique.
In the following, we omit this particular $F$ in the naming since the final constructions are independent thereof.
For \eqref{eqn:NI}, a particular marginal function reads \cite{Kimura2013-control}:
\begin{equation} \label{eqn:NI-F}
	F(x, \theta) := x_1^4 + x_2^4 + \frac{\abs{x_3}^3}{(x_1 \cos (\theta) + x_2 \sin(\theta) + \sqrt{\abs{x_3}})^2}.
\end{equation}
It can be shown that \eqref{eqn:NI} can be practically stabilized by a control law
\begin{equation}
	\label{eqn:NI-ctrl}
	\kappa(x; \theta^\star) = -\begin{pmatrix}
		\scal{\zeta(x; \theta^\star), g_1(x)} \\ \scal{\zeta(x; \theta^\star), g_2(x)}
	\end{pmatrix}.
\end{equation}
where $\zeta$ is a disassembled subgradient with a corresponding minimizer $\theta^*$.
The work \cite{Kimura2013-control,Kimura2015-asymptotic} suggests to pick the following disassembled subgradient:
\begin{equation}
	\label{eqn:NI-analyt-subgrad}
		\zeta_a(x) := \begin{cases}
			\nabla V(x), & (x_1, x_2) \neq (0,0), \\
			\pdiff{F(x, 0)}{x}, & (x_1, x_2) = (0,0).	
		\end{cases}
\end{equation}
We use subscript ``a'' here to indicate that this disassembled subgradient is computed analytically (roughly speaking).
In contrast to it, we may take \eqref{eqn:disass-diff} for \eqref{eqn:NI} using the marginal function \eqref{eqn:NI-F} directly and extract a selector, call it $\zeta_s$, by Algorithm \ref{alg:meas-sel-extraction}.
The respective selector was extracted for $eps = \nicefrac{1}{16}$ and the computation took about half an hour in MATLAB 2019 on a 3.4 GHz/ 8 GB RAM machine.
Fig. \ref{fig:NI-sim} shows, as an example, how a section of the selector-based disassembled subgradient $\zeta_s$ looks like for $x_3 = -1$.

\begin{figure}[h]	
	\centering
	\includegraphics[width=0.45\textwidth]{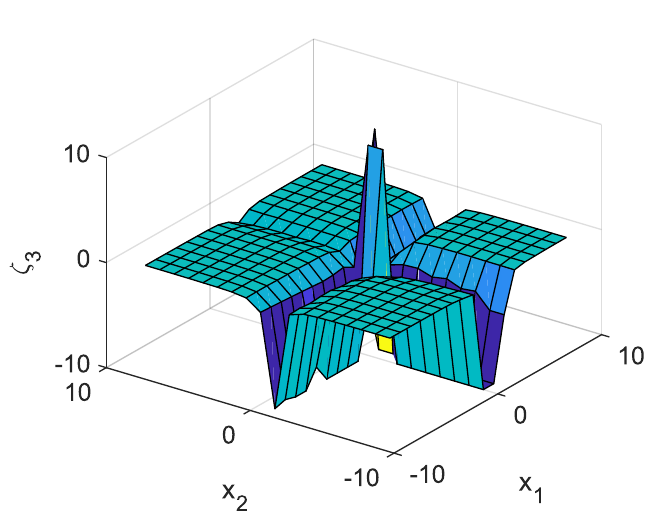}
	\caption{Exemplary section of the extracted selector, the value of the disassembled subgradient $\zeta_3$ at $x_3=-1$}
	\label{fig:NI-sim}
\end{figure}

This was, however, an offline part.
In the actual simulation, we only needed to evaluate the selector to compute control actions by $\kappa$ \eqref{eqn:NI-ctrl}.
The controller was sampled at 10 ms, which is reasonable for industrial micro-controllers.
The simulation of 10 s took less than a minute.
The results are displayed in Fig. \ref{fig:NI-sim}.
What is interesting is that the $\zeta_a$-based controller is much more prone to numerical uncertainty than the selector-based one.
We suspect this might be due to the case distinction in \eqref{eqn:NI-analyt-subgrad} (refer also to the discussion in Example \ref{exm:count-red}).
As a result, although seemingly a bit slower, the convergence of the state in case of $\zeta_s$ happens to have almost no chattering at all.

To conclude this section, it should be remarked that the current computational study was dedicated to the principal realizability of computationally effective measurable selector extraction, and no general claim, that selector-based controls yield better stabilization results, can be made at the moment.
It requires an analysis going beyond the current work, but the latter provides at least some insights into computational aspects of Theorem \ref{thm:meas-sel}.
It should be decided in each particular application what has the priority: verified computation and safety or computation speed.
At least for the former scenario, the current work proposes a machinery, which may be useful.

\begin{figure*}[h]	
	\centering
	\includegraphics[width=0.45\textwidth]{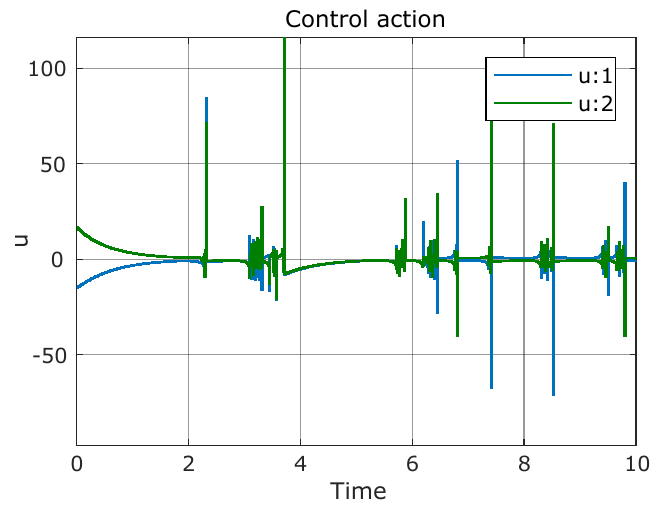}
	\includegraphics[width=0.45\textwidth]{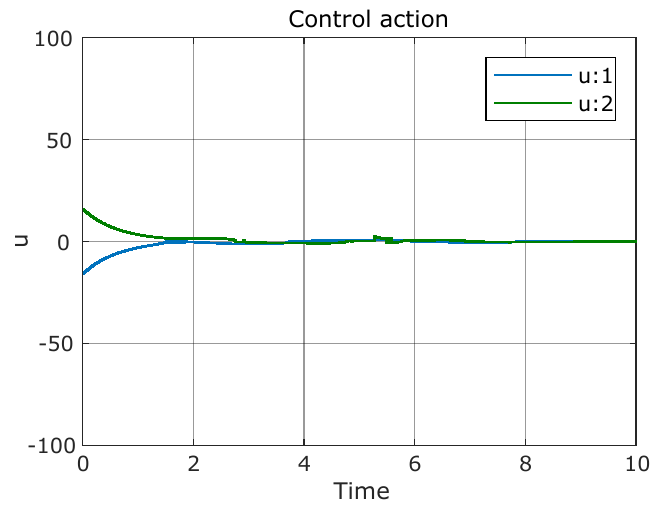}
	\includegraphics[width=0.45\textwidth]{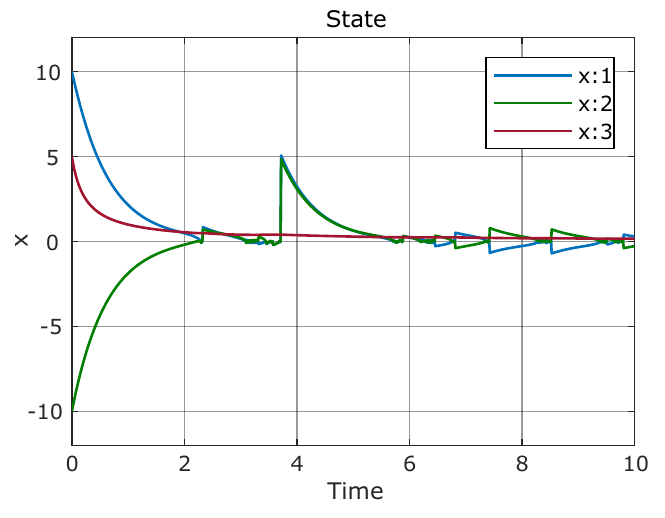}
	\includegraphics[width=0.45\textwidth]{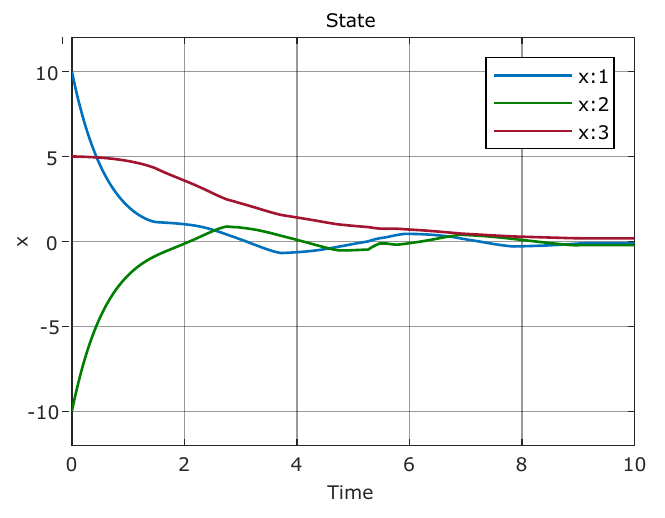}	
	\caption{Control and state of the three-wheel robot in a simulation with a practically stabilizing controller that uses: the analytic disassembled subgradients $\zeta_a$ of \eqref{eqn:NI-analyt-subgrad} (left); selector-based disassembled subgradients $\zeta_s$ (right)}
	\label{fig:NI-sim}
\end{figure*}

\section{Conclusion}

This work was dedicated to constructive extractability of measurable selectors widely used in control applications.
Available theorems do not allow constructive extraction of selectors, as was pointed out by \cite{Aubin2011-viab-thr} and demonstrated in a counter-example in Section \ref{sec:intro}.
A constructive version of the measurable selector theorem was proven.
To this end, the property called ``representability'' of domains and SVFs was introduced and analyzed.
A variant of the theorem was also proven for representable SVFs satisfying a certain weak continuity condition.
Some selected applications, including viability theory, differential inclusions and optimal control were discussed.
A computational study with a three-wheel robot model practically stabilized by a selector-based controller was given.


\bibliographystyle{IEEEtran}
\bibliography{bib/constr-math,bib/Osinenko,bib/set-thr,bib/analysis,bib/viability,bib/nonlin-ctrl,bib/diff-games,bib/opt-ctrl,bib/topology,bib/computable,bib/TAC,bib/selector-misc,bib/Filippov-sol,bib/sliding-mode,bib/non-smooth-analysis,bib/DP,bib/stabilization,bib/semiconcave,bib/form-ver-ctrl}


\section*{Appendix}

\begin{prp}
	For any generalized interval $J = \bigcup_i I_i$, its countable reduction $K$ is dense in $J$.  
	\label{prp:count-red-dense}
\end{prp}
\noindent For the proof of Proposition \ref{prp:count-red-dense}, we utilize the following fact:
\begin{fct}
	Let $A, B$ be any finite unions of basic intervals. Let $A'$ be dense in $A$. Then, $A' \bigcup B \sm A$ is dense in $A \bigcup B$.  
	\label{fct:count-red-dense-simple}
\end{fct}
\begin{prf}
Let $x, \eps > 0$ be arbitrary. We find a $y$ in $A' \bigcup B \sm A$ \sut $|x-y| \le \eps$. If $x \in A$, we set $y := x$. If $x \in B$, then consider the two cases: $|B - A| > 0 \lor |B - A| < \frac{\eps}{2}$. In the former case, $B \sm A = B$ and we set $y := x$. In the latter case, it holds that $|x - A \bigcap B| > 0 \lor |x - A \bigcap B| < \frac{\eps}{2}$. In the former case, $x \in B \sm A$, and we set $y := x$. In the latter case, there is a $z \in A \bigcap B$ \sut $|x - z| \le \frac{\eps}{3}$. But since, trivially, $|A - A \bigcap B| \le \eps$, it holds that, also, $|A - z| < \frac{\eps}{3}$, whence there is a $w \in A$ with $|w-z| < \frac{2 \eps}{3}$. Since $A'$ is dense in $A$, there is a $y \in A'$ \sut $|x-y|<\eps$. 
\end{prf}
\noindent Proceed to the proof of Proposition \ref{prp:count-red-dense}:

\begin{prf}
The proof is done by induction as follows: assume $\bigcup_{k \le i} \left( I_k \sm \bigcup_{l<k} I_l \right)$ is dense in $\bigcup_{k \le i} I_k$. We show that $\bigcup_{k \le i+1} \left( I_k \sm \bigcup_{l<k} I_l \right)$ is dense in $\bigcup_{k \le i+1} I_k$. If $i=2$, the result holds by Fact \ref{fct:count-red-dense-simple}. Denote
\begin{align*}
A' := \bigcup_{k \le i} \left( I_k \sm \bigcup_{l<k} I_l \right), & & A := \bigcup_{k \le i} I_k, & & B := I_{i+1}.
\end{align*}
Then, $\bigcup_{k \le i+1} \left( I_k \sm \bigcup_{l<k} I_l \right) = A' \bigcup B \sm A$ and $\bigcup_{k \le i+1} I_k = A \bigcup B$. By the induction hypothesis and Fact \ref{fct:count-red-dense-simple}, the result follows.
\end{prf}
\begin{crl}
	The countable reduction $K = \{K_n\}_n$ of any sequence of generalized intervals $X = \{J_n\}_n$ is dense in $X$.
\end{crl}

\begin{thm}[Filippov's solution to Lipschitzean DI\cite{Aubin2012-diff-incl,smirnov2002introduction}]
	Consider an interval $\T = [0, T]$, an absolutely continuous function $g: \T \ra \Y$, and let $A \subseteq \T \times \Y$ with $(t,x) \in A \implies \in | x - g(t)| \le \beta$ for some $\beta > 0$.
	Let $F: A \rra \Y$ be an SVF with closed values and satisfying a Lipschitz condition $|F(t,x) - F(t, y)| \le \kappa(t) |x - y|$ (in the sense of Hausdorff metric) with $\kappa$ non-negative, integrable on $\T$..
	Assume that $|x_0 - g(a)| = \delta \le \beta, |\dot g(t), F(t, g(t))| \le p(t) \text{ a.~e.}$
	with $p \in L^1(\T)$. Set
	\[
	\xi(t) := \delta e^{\int \limits_a^t \kappa(\tau) \diff \tau} + \int \limits_a^t e^{\int \limits_{\tau}^t \kappa(s) \diff s} p(\tau) \diff \tau
	\]
	and let $J := [a, \omega]$ be a non-empty interval with $t \in J \implies \xi(t) \le \beta$. Then, there exists a solution $x(t)$ on $J$ to the problem
	\[
	\dot x \in F(t, x(t)), x(a) = x_0
	\]
	\sut
	\begin{align*}
	& | x(t) - g(t) | \le \xi(t) & \\
	& | \dot x(t) - \dot g(t) | \le \kappa(t) \xi(t) + \rho(t), & \text{a.~e.}
	\end{align*}
	\label{thm:Filippov-sol-exist-Lip-DI}
\end{thm}

\begin{prf} \textbf{(Idea)}
The proof of Theorem \ref{thm:Filippov-sol-exist-Lip-DI} constructs a sequence of absolutely continuous functions $\{x_n(\bullet)\}_n$, then shows that it is Cauchy and, thus, converges uniformly to a function $x(\bullet)$ which is a required solution.
At each step $n$, a measurable selector $v_n$ is extracted for $F(\bullet, x_n( \bullet ))$ satisfying
\[
x_{n+1}(t) = x_0 + \int_a^t v_n(\tau) \diff \tau
\]
and it is shown that $\{v_n\}_n$ converges almost everywhere, whence 
\[
x(t) = x_0 + \int_a^t v(\tau) \diff \tau
\]
and $v(\tau) \in F(\tau, x(\tau))$ a.~e. on $J$. Crucial for the proof is the extraction of measurable selectors $v_n, n \in \N$.
Therefore, if $F(\bullet, x_n( \bullet ))$ are representable, Theorem \ref{thm:meas-sel} can be used to find a solution in the constructive setting.
\end{prf}


\end{document}